\documentclass[aps,reprint,pre,twocolumn,showpacs,groupedaddress]{revtex4}
\usepackage{graphicx}
\usepackage{bm}
\usepackage{comment} 
\usepackage{amssymb}
\usepackage{color}
\usepackage{amsbsy}      
\usepackage{amsmath}
\usepackage{slashbox}
\def\dd{\mbox{d}}

\def\f{{\bf f}}
\def\r{{\bf r}}

\def\k{{\bf k}}
\def\u{{\bf u}}
\def\v{{\bf v}}
\def\x{{\bf x}}

\def\F{{\bf F}}

\def\R{{\bf R}}

\begin{document}

\title{Swarming in viscous fluids: three-dimensional patterns in
  swimmer- and force-induced flows}

\author{Yao-Li Chuang} 

\affiliation{Dept.~of Mathematics, California State University,
  Northridge, Northridge, CA 91330} 

\author{Tom Chou} 
\affiliation{Depts. of Biomathematics \&
  Mathematics, UCLA, CA 90095-1766}

\author{Maria R. D'Orsogna}
\affiliation{Dept. of Mathematics, California State University,
  Northridge, Northridge, CA 91330} 

%\maketitle   %comment out for revtex

%\runninglinenumbers*

%\begin{article} %comment out for revtex

\begin{abstract} % 
\noindent
We derive from first principles a three-dimensional theory of
self-propelled particle swarming in a viscous fluid environment.
Our model predicts emergent collective behavior that depends
critically on fluid opacity, mechanism of self-propulsion, and type of
particle-particle interaction. In ``clear fluids" swimmers have full
knowledge of their surroundings and can adjust their velocities with
respect to the lab frame, while in ``opaque fluids," they control
their velocities only in relation to the local fluid flow.
We also show that ``social'' interactions that affect only a
particle's propensity to swim towards or away from neighbors induces a
flow field that is qualitatively different from the long-ranged flow
fields generated by direct ``physical'' interactions. The latter can
be short-ranged but lead to much longer-ranged fluid-mediated
hydrodynamic forces, effectively amplifying the range over which
particles interact.
These different fluid flows conspire to profoundly affect swarm
morphology, kinetically stabilizing or destabilizing swarm
configurations that would arise in the absence of fluid. Depending
upon the overall interaction potential, the mechanism of swimming
({\it e.g.}, pushers or pullers), and the degree of fluid opaqueness,
we discover a number of new collective three-dimensional patterns
including flocks with prolate or oblate shapes, recirculating
peloton-like structures, and jet-like fluid flows that entrain
particles mediating their escape from the center of mill-like
structures. Our results reveal how the interplay among general
physical elements influence fluid-mediated interactions and the
self-organization, mobility, and stability of new three-dimensional
swarms and suggest how they might be used to kinetically control their
collective behavior.
\end{abstract}

\maketitle

\noindent
The collective behavior of self-propelled agents in natural and
artificial systems has been extensively studied
\cite{SPO91,KAU93,VIC95,TON95,ALB96,FLI99b,SHIMOYAMA96,EDE98,LEV00,EBE01,PAR02,GRE03,CHA04,COU05,DOL05,MOR05,DOR06,CHA06,CHU07b,BER13,KAI13,YOU13}.
Many of the lessons learned from experimental and theoretical work
conducted on organisms as diverse as bacteria, ants, locusts, and
birds
\cite{SCH71,KIM90,NIW94,DIL95,ROM96,KOC98,PAR99,CAM03,ORD03,OIE04,GOM05,MAC06,TOP08,HEI13}
have been successfully applied to engineered robotic systems to help
frame decentralized control strategies through ad-hoc algorithms
\cite{SUG97,BON99,LEO01,GAZ03,CHA03,JAD03,LIU03,CHU07a}.
In most mathematical ``swarming'' models, particles are assumed to be
self-driven by internal mechanisms that impart a characteristic speed.
A pairwise short-ranged repulsion and a long-ranged decaying
attraction are typically employed as the most realistic choices when
modeling aggregating particles
\cite{LEV00,PAR02,COU02a}.  The interplay between self-propulsion,
particle interactions, initial conditions, and number of particles is
key in determining the large scale patterns that dynamically arise.
In two dimensions, rotating mills and translating flocks are often
observed, the latter configuration also arising in three dimensions
\cite{LEV00,EBE01,DOR06,CHU07b,STR08,NGU12}. It is possible to
classify swarm morphology in terms of interaction strength and length
scales, as shown for particles coupled via conserved forces derived
from the Morse potential \cite{DOR06,CHU07b}. 
Externally applied potentials and noise can be also used to trigger
transitions between coherent and disordered structures
\cite{VIC95,STR08,NGU12}.

Although different rules for the characteristic speed have been
proposed \cite{LEV00,EBE01,DOR06}, most studies so far have focused on
self-propelled agents in ``vacuum'', ignoring the medium in which
nearly all real systems operate. One exception is the literature on
swimmers wherein models have been developed for a single or a few
organisms that propel themselves in viscous
\cite{TAY51,HAN53,BLA73,SHA89,GAL99,BEC03,BAS08,SPA12,WOL08,LAU09a,HUB11} and
non-Newtonian fluids
\cite{ROS74,CHA79,FUL98,LAU07,NOR08,LAU09a,LAU09b,FU07,FU09,ZHU12}.  
In particular, swarming hydrodynamic theories have been derived wherein
swimmer densities with or without fluid flows are described as continuous
fields \cite{BAS09, TON95,CHU07b,SOL15}. These
%
%modeled as continuous fields \cite{BAS09}. However, these 
%
%{\color{red} coarse-grained\/}
%
``two fluid" models however may not always
display the rich features observed when particles retain their
discreteness, especially in terms of finite-sized swarm morphology,
stability and self-organization.

To efficiently study the collective dynamics of self-propelled
particles in a fluid medium we derive a microscopic three-dimensional
``agent-based'' kinetic theory that incorporates hydrodynamic
interactions between particles.
A possible starting point would be to assume that the fluid medium
leads to direct coupling between particle velocities and build this
effect into existing models using, for example, the Cucker-Smale
velocity matching mechanism. Here, particle $i$ is subject to an
additional force due to the presence of particle $j$, given by $\v_j -
\v_i$ and modulated by a distance-dependent prefactor $g(\vert
\r_{i}-\r_{j}\vert)$ \cite{CUC07a,CUC07b,SHE08}. Because of its
simple mathematical form, Cucker-Smale type interactions have been
used extensively to study swarming, with coherent morphologies arising
depending on the form of $g(\vert \r_{i}-\r_{j}\vert)$. Although not
explicitly meant to model fluid-mediated couplings, a heuristic
Cucker-Smale-type interaction could be constructed by choosing an
appropriate form for $g(\vert \r_{i}-\r_{j}\vert)$ or different powers
of $|\v_i - \v_j|$.  Whether such an approach would be consistent with
more microscopic derivation of fluid-mediated
particle-particle interactions is however unclear.

The goal of this paper is to derive from first principles a theory of
particle swarming in fluids and to understand the ways viscous flows
can arise and affect particle dynamics and collective behavior.  In
order to incorporate fluid couplings into discrete particle models,
one must first identify the physical origin of the interactions
between particles. In typical models of swarming
\cite{LEV00,GRE03,MOR05,DOR06,TOP08,BER13}, the propensity of agents
to self-propel themselves towards or away from others is modulated by
an effective ``social'' interaction potential.  When immersed in
low-Reynolds number Newtonian fluids, particle self-propulsion is
force-free. Here, the fluid flows arising from swimming or squirming
particles can be decomposed in terms of force dipoles or higher order
force distributions leading to velocities that decay away from the
swimmer as $1/r^{n}, n\geq 2$
\cite{NAS97,COR04,COR05,RIE05,ISH06,LAU09a}. The sign and amplitude of
this self-propulsion-induced flow field depend on the specific details
of the ``stroke'' of the swimmer \cite{NAJ04,WOL08,BAS09,HUB11,SPA12}.

A qualitatively different flow arises if the potential is associated
with a true ``physical'' force, arising from, say, electrostatic
molecular, magnetic, or gravitational interactions. These physical
interactions between swimmers impart an external body force on each of
them, ultimately leading to a flow field that decays as $1/r$
\cite{HAP65,ESP95,BATCHELOR}.  Although the physical forces between
particles may be short-ranged, they can be transmitted to the
surrounding fluid \cite{OSE27,LAD63,FIN65,BLA71,ESP95,SHU01},
collectively generating a much longer-ranged flow field, and
effectively \textit{extending} the range of interactions. Thus, the
resulting $1/r$ Oseen flow field can be even longer ranged than the
$1/r^{n}$ flows arising from self-propulsion.

The different origins of fluid flow can be most easily understood by
considering a single particle moving under a constant chemical
gradient or passively sedimenting under a gravitational potential. A
chemoattractant can be represented by a ``social'' interaction as it
only directs a force-free self-propeller towards a particular
velocity.  A body force resulting from \textit{e.g.}, gravity is a
physical force since it ultimately imparts a force on the fluid.  In
both cases, particle trajectories are identical and can be described
by motion under a linear effective potential. However, within a fluid
medium, a chemotactic social interaction generates a different flow
from that of sedimentation under a physical interaction. As we shall
see, the qualitatively different fluid flows arising from social and
physical interactions also lead to qualitatively different collective
particle behavior. How the details of particle-particle interactions
are modeled and interpreted thus becomes a critical element in the
development and application of hydrodynamically coupled particle swarming
theories.

Finally, we also consider two different types of fluids: ``clear'' and
``opaque.''  In a clear fluid, particles can ``see'' fixed markers and
have direct knowledge of their motion in reference to the rest
frame. Their absolute velocities can be directly controlled by their
internal self-propelling mechanism.  Here, the surrounding fluid
simply imparts an additional drag force. In the richer and more
interesting case of an opaque fluid, particles only have near-field
vision and their velocities can be governed only in relation to the
surrounding flow.  For both social and physical interactions, we systematically
derive the effective particle-particle coupling arising from viscous
Stokes flows and investigate their effects on coherent three
dimensional swarming structures.  These hydrodynamic interactions
strongly affect collective dynamics and give rise to surprising new
patterns such as distorted flocks, pelotons, core-filled mills, and
mills that perpetually disband and reform.

\section*{Fluid-coupled equations of motion}

\noindent

\noindent
We consider a system of $N$ identical particles
with mass $m=1$. We can write the equations of motion for
particle $i$  at position 
$\r_{i}(t)$ and lab-frame velocity $\v_{i}(t)$ as follows

\begin{eqnarray}
\nonumber
\dot{\r}_{i}&=& \v_{i}, \\
\dot{\v}_{i}  &=&   - \gamma(\v_i - \u(\r_i))
+ \f_M(\v_i, \u(r_i)) + \f_i
%\left( \alpha - \beta \left| \v_{i} - 
%\lambda \u(\r_{i}) \right|^2 \right) \left( \v_{i} - \lambda \u(\r_{i}) 
%\right)  \nonumber \\
%\: & \hspace{3mm} -\gamma (\v_{i} - \u(\r_{i})) - \nabla_i \sum_{j\neq i} 
%\Phi(\vert \r_{i}-\r_{j}\vert).
\label{MDOT3}
\end{eqnarray}

\noindent 
where $\u(\r_{i})$ is the lab-frame fluid velocity generated at position
$\r_{i}$ by the motion of all other particles in the absence of
particle $i$. 
In Eq.\,\ref{MDOT3} the drag force $-\gamma (\v_i - \u(\r_i))$ on particle
$i$ is proportional to its velocity relative to that of the
fluid $\u(\r_i)$. Without loss of generality, we assume spherical particles
with a small radius $a$ and drag coefficient $\gamma =6\pi \eta a$.
The force term $\f_{\rm M}(\v_{i}, \u(\r_i))$ represents the self-propelling
motility force on particle $i$, which can depend both on $\v_i$ and
$\u(\r_i)$.  
%Here $\u(\r_{i})$ is the lab-frame fluid velocity generated at position
%$\r_{i}$ by the motion of all other particles in the absence of
%particle $i$. 
%The first term on the right-hand side of Eq.~\ref{MDOT3}
%is a proxy for self-propulsion, the second term is the drag force,
%and the third term describes direct social or physical interaction 
%force acting on particle $i$ by the other particles.
%The force term $\f_M(\v_i, \u(r_i))$ is a proxy for self-propulsion
%and can depend both on $\v_i$ and on $\u(\r_i)$. It is  
%Ignoring couplings to the surrounding medium and drag forces, $\f_M$
It is usually chosen to have non-trivial zeros that identify characteristic speeds 
of motion.  For the purposes of this paper, we adopt a
modified friction form given by

\begin{eqnarray}
\label{fric}
\f_M(\v_i, \u(\r_i)) = (\alpha - \beta |\v_i - \lambda \u (\r_i)|^2),
\end{eqnarray}

\noindent
where the parameters $\alpha$ and $\beta$ quantify self acceleration
and deceleration, respectively.  Eq.\,\ref{fric} with $\lambda=0$ yields
the classical Rayleigh-Helmholtz friction that has
 been extensively used to model self-propulsion in vacuum
 \cite{DOR06,CHU07b,ROM12,NGU12}.  In this case, the natural
characteristic speed arises by setting $\f_M$ to zero and is given by $|\v_i| =
\sqrt{\alpha /\beta}$.  We introduce the ``perception
coefficient'' $\lambda$ in Eq.\,\ref{fric} to represent how well particles sense their
environment once they are immersed in a fluid. 
%This parameter will indicate whether the fluid is clear or opaque.
%Setting the term to zero
%leads to the characteristic speed $\left| \v_i - \lambda \u (\r_{i}) \right|
%\approx\sqrt{\alpha/\beta}$, where $0 \le \lambda \le 1$ is named 
%a ``perception coefficient.''
The choice $\lambda = 0$ represents the case where swimmers can
determine their lab frame velocities $\v_i$ as if they were in a vacuum
and adjust their speed towards the characteristic velocity $\sqrt{\alpha/ \beta}$.
 Thus $\lambda=0$ indicates a
``clear" fluid where any effects on particle movement imparted to particles by the fluid will arise only
through drag forces.  
Conversely, the choice $\lambda = 1$ indicates that swimmers 
have no knowledge of the lab frame and can determine their motion only in
relation to the local fluid.  As a result, swimmers will regulate their relative velocity $\v_{i} -
\u(\r_{i})$ and not their lab frame velocity $\v_i$ toward the characteristic speed. 
This is the ``opaque" fluid limit.  Other values
$0 < \lambda < 1$ yield intermediate regimes. Note that the
Rayleigh-Helmholtz friction is not the only option for modeling
swarming self propulsion. An in-depth discussion about the effects of
choosing different functional forms for self propulsion can be found
in \cite{ROM12}; the difference is particularly profound in the
presence of noise. Finally

\begin{equation}
   \f_i = - \nabla_{i} \sum_{j\neq i} \Phi (\vert
     \r_{i}-\r_{j}\vert)
   \label{eq:int_forcei}
\end{equation}

\noindent
is the particle-particle interaction force on particle $i$, where
$\nabla_{i}\equiv \partial/\partial \r_{i}$ and $\Phi (\vert
\r_{i}-\r_{j}\vert)$ is the direct pairwise interaction potential.   
While any mathematical form for $\Phi(\vert (\r_i - \r_j) \vert$ can be used, to be
consistent and comparative with previous literature, we use
the commonly studied Morse potential \cite{LEV00}, given by the
superposition of repulsive and attractive components

\begin{equation}
   \Phi \left( \left| \r_{i}-\r_{j} \right| \right) = 
            C_r \mathrm{e}^{- \frac{\left| \r_{i}-\r_{j} \right|}{\ell_r}} 
           - C_a \mathrm{e}^{-\frac{\left| \r_{i}-\r_{j} \right|}{\ell_a}}.
            \label{eq:MorsePotential}
\end{equation}

\noindent
The coefficients $C_a$ and $C_r$ in Eq.~\ref{eq:MorsePotential} define
the strengths of the attractive and repulsive potentials, respectively
and $\ell_a$ and $\ell_r$ specify their effective lengths of
interaction. Using this potential, the fluid-free swarming problem
(Eq.~\ref{MDOT3} with $\u = 0$) has been very well studied especially in
one and two dimensions
\cite{LEV00,DOR06,CHU07b,BER13,TOP04}. 
Generally, particles are
subject to two tendencies: changing their separations to minimize
$\Phi$ and adjusting their velocities to match the characteristic speed. 
Depending on initial conditions, dimensionality, number of
particles and/or parameter choices, both tendencies can be
simultaneously satisfied, leading, for example, to rigidly translating
flocks. If only one is satisfied, mills, rigid disks, or random motion
arise \cite{CHU07b}.

For the fluid-coupled equations of motion what now
remains is to specify the source of $\u$.  To model $\u$ we note that
in classical swarming models the potential $\Phi$ is usually a
mathematical representation of socially derived interactions. In this
scenario, the only actual force exerted by the swimming particles on
the fluid is via their self propulsion.  In this case, the fluid
disturbance $\u \equiv \u_{\rm s}$ depends only on the microscopic
details of the swimming mechanism and decays as $1/r^{n}, n\geq 2$
\cite{NAS97,COR04,COR05,RIE05,ISH06,LAU09a}.  In this paper we assume
an effective stroke-averaged self-propulsion whereby the swimmer's
period-averaged strokes are described as a force dipole acting on the
fluid leading to

\begin{equation}
   \u_{\rm s}(\r) = 
      \sum_{j} \frac{G v_{j}}{R_{j}^2}
      \left[ 3 \left( \hat{\R}_{j} \cdot \hat{\bf v}_j \right)^2 - 1 \right]
      \hat{\R}_{j},
   \label{eq:uv}
\end{equation}

\noindent
where $\R_j \equiv (\r - \r_j)$, $R_j \equiv \left| \R_j \right|$,
$\hat{\R}_j \equiv \R_j / R_j$, $v_j \equiv \left| \v_j \right|$, and
$\hat{\v}_j \equiv \v_j / v_j$ \cite{BAS09}.  Here, the social
potential $\Phi$ influences $\u_{\rm s}(\r)$ through $v_{j}$ which
obeys Eq.\,\ref{MDOT3}.  The lumped parameter $G$ in Eq.\,\ref{eq:uv}
depends on the details of swimmer geometry such as its length and
longitudinal mass distribution, and carries units of a length squared.
For $G>0$, the orientation of the force dipole is parallel to the
swimmer's velocity, describing a propelling swimmer or a ``pusher'';
conversely, $G<0$ denotes a contractile swimmer or a ``puller.'' This
pusher/puller model of self-propulsion in viscous Stokes flow has been
often used in models of swimmers.

Now, if a true action-at-a-distance physical force arises between
particles, the latter will experience body forces during the course of their
dynamics.  In addition to a self-generated flow field $\u_{\rm s}(\r)$,
such particles will transfer their body force to the fluid resulting
in an additional flow field $\u_{\rm p}(\r)$.  
For incompressible low-Reynolds number fluids, we can find
$\u_{\rm p}(\r)$ by solving, in the quasi-static limit, Stokes' equation
with an added interaction-mediated force density $\F \left(
\r \right) \approx -\sum_i \sum_{j\neq i} \delta \left(\r - \r_i
\right) \nabla_{i} \Phi (\r_{i}-\r_{j})$:

\begin{equation}
      \rho \displaystyle{\frac{\partial \u_{\rm p} }{\partial t}} 
 = \eta \nabla^2 \u_{\rm p} - \nabla p + \F(\r).
  \label{eq:Stokes_tdep}
\end{equation}
Here, $\rho$ and $\eta$ are the density and the dynamic viscosity of
the fluid, $p$ is the local pressure, and $\delta \left(\x\right)$ is
the Dirac delta-function. In three dimensions, the analytic solution
to $\u_{\rm p}(\r)$ is expressed in terms of the static Oseen tensor

\begin{equation}
\u_{\rm p}(\r) = - \sum_{j}\sum_{k \neq j}
\frac{\left[{\bf I} +  \hat{\R}_j \hat{\R}_j \right]} {8\pi \eta
    R_j} \cdot \nabla_j \Phi(\vert \r_{j}-\r_{k} \vert ),
\label{USTATIC}
\end{equation}

\noindent where ${\bf I}$ is the identity matrix.  Note that in
contrast to the $1/R^{2}$ dependence of $\u_{\rm s}(\r)$ in
Eq.~\ref{eq:uv}, $\u_{\rm p}(\r)$ is longer ranged, decaying as
$1/R$. Also, note that while we neglect the inertia of the fluid in
Eq.\,\ref{eq:Stokes_tdep} we retain particle inertia in
Eq.\,\ref{MDOT3}, implicitly assuming that particle mass density is
much higher than fluid mass density.

In general, self-propulsion in a Stokes fluid will only generate $\u_{\rm
  s}(\r)$. The longer-ranged flow $\u_{\rm p}(\r)$ arises only if the
interaction potential is associated with a physical interaction that
imparts a body force on particles and fluid.  In this case, the linearity of
the Stokes fluid dynamics allows us to decompose 
$\u(\r) \equiv \u_{\rm s}(\r) + \chi \u_{\rm p}(\r)$, where
$\u_{\rm s}(\r)$ and $\u_{\rm p}(\r)$ are given by Eqs.\,\ref{eq:uv}
and \ref{USTATIC}, respectively. To separate the effects of the
different flow fields, we introduce the toggle $\chi=0$ or 1 in
the definition of $\u$ which allows us to switch off the physical
force-induced flow field $\u_{\rm p}$ (by setting $\chi =0$). To
switch off swimmer-induced flows $\u_{\rm s}$ we can set $G=0$ and $\chi =1$.  The
inclusion of both flows requires a non-zero $G$ and $\chi = 1$.

In the remainder of this paper, we investigate swarming coupled to
viscous Stokes flows.  Inertial flows given by the complete
time-dependent solution of Eq.~\ref{eq:Stokes_tdep} can be expressed
in terms of a dynamic Oseen tensor as shown in the Appendix.  In the extreme
limit of $\nu \to 0$, either the fluid inertia is too large ($\rho \to
\infty$) to induce any flow field, or the fluid becomes inviscid
($\eta \to 0$). For inviscid fluids the induced hydrodynamic
interaction can be described as a potential flow and is dipolar, which
is even shorter-ranged than the force-dipole-generated $\u_{\rm
  s}(\r)$ considered in this paper. For completeness, we derive
interaction-induced inviscid fluid flow also in the Appendix.

Henceforth, we non-dimensionalize space and time according to
$\r^\prime = \frac{\sqrt{\alpha \beta}}{m} \r$ and $t^\prime =
\frac{\alpha}{m} t$. All other dimensionless model parameters are
given in the Appendix. We also drop the prime superscripts and
define the full fluid-coupled swarming model as $\dot{\r}_i = \v_{i}(t)$
and

\begin{align}
  \dot{\v}_{i} = & \left(1 - \left| \v_{i} - \lambda \u(\r_{i}) \right|^2 \right)
\left( \v_{i} - \lambda \u(\r_{i}) \right) \nonumber \\
\: & \hspace{9mm} - \gamma(\v_{i}-\u(\r_{i}))- 
\nabla_i \sum_{j\neq i} \Phi(\vert\r_{i}-\r_{j}\vert).
\label{MDOT4}
\end{align}

\noindent
We numerically solve Eq.~\ref{MDOT4} with $\u_{\rm s}(\r)$ given by
Eq.~\ref{eq:uv} and $\u_{\rm p}(\r)$ given by Eq.~\ref{USTATIC} using the
fourth-order Runge-Kutta method with an adaptable time-step \cite{GOL87}. 
Since both $\u_{\rm s}(\r)$ and $\u_{\rm p}(\r)$ depend on
particle positions, they are updated at each time step. Initial
conditions are defined by still particles placed at uniformly
distributed random positions within a $3 \ell_a^3$ box which is
removed after the start of the simulation.  Unless otherwise
specified, we set dimensionless Morse-potential parameters to $C_a =
1.0$, $\ell_a = 2.0$, $C_r = 2.0$, $\ell_r = 1.0$, representing
long-ranged attraction and short-ranged repulsion
\cite{EDE98,LEV00,CHU07b}. The effects of varying potential parameters
$C_{r,a}$ and $\ell_{r,a}$ are discussed in the Appendix. 

\begin{comment}
While the systematic parameter studies presented in this paper are based on simulations 
of $50$ particles, we have conducted preliminary simulations up to 
$5000$ particles and found no qualitative differences in terms of
swarming morphologies. This is 
consistent with the two-dimensional fluid-free model with the potential 
parameters in the explored regime here (known as the catastrophic potential 
regime) \cite{CHU07b}.
\end{comment}

Finally, to counter the collapsing tendency between particle pairs 
due to the $1/R^{2}$ dependence of $\u_{\rm s}(\r)$, we add to 
$\Phi(|\r_i - \r_j|)$ an extremely short-ranged diverging repulsive 
potential $\sim 1/R^{12}$ to keep particles reasonably apart. 
We numerically investigate our model for different values of dynamic 
viscosity $\eta$ and swimmer propulsion strength $G$.

%\vspace{2mm}

\section*{Results and Discussion}

\vspace{1mm}

\begin{figure}[ht]
  \begin{center}
      \includegraphics[width=2.8in]{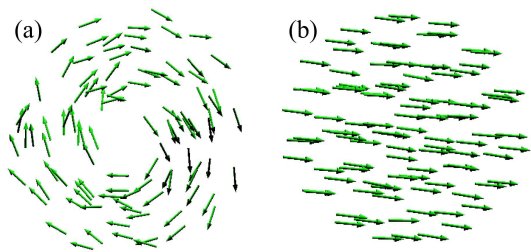} 
\end{center}
  \caption{Snapshots of typical three-dimensional swarm patterns for
    100 particles. (a) a rotational mill, and (b) a translating
    flock.}
\label{fig:snapshots}
\end{figure}

\noindent \textit{\textbf{Fluid-free limit -- }}For reference, we
first consider $\lambda=\gamma=0$ where particle and fluid dynamics
decouple. Eqs.~\ref{MDOT3} and \ref{eq:MorsePotential} now reduce to the
three-dimensional version of the well-studied two-dimensional swarming
model presented in \cite{DOR06,CHU07b}.  While studies of
three-dimensional swarms have previously been examined \cite{NGU12},
the full dynamics including the emergence of transient structures have
not been explored.  For the interaction parameters chosen above,
possible coherent states in two-dimensional include a flock, a single
rotating mill, and two counter-rotating mills \cite{CAR09}.  In three
dimensions we do not find counter-rotating mills: only simple mills
and spherically shaped flocks can arise from random initial conditions, 
as shown in
Fig.~\ref{fig:snapshots}. The absence of counter-rotating mills in
three dimensions can be easily understood.  In two dimensions there
are only two possible rotating directions, but in three dimensions
there are an infinite number of rotational axes.  Reversing rotating
directions in two dimensions requires the angular momentum to change
signs, but in three dimensions the rotational axes of a sub-mill can
continuously evolve along the third dimension until all particles
eventually come to rotate about a common axis.  This picture is
consistent with diffusion of angular momentum in three-dimensional
swarms \cite{STR08}.

%%%%%%%%%%%%%% Snapshot Images %%%%%%%%%%%%%%%%%%%%%%%%%%%%%%%%%%%%%%%%%%%%%%%
%%%%%%%%%%%%%%%%%%%%%%%%%%%%%%%%%%%%%%%%%%%%%%%%%%%%%%%%%%%%%%%%%%%%%%%%%%%%%
Most notably, despite being the dominant steady state in two
dimensions, the single rotating mill shown in
Fig.~\ref{fig:snapshots}(a) is not a true three-dimensional steady
state. Although particles may settle into identifiable mills,
extensive simulations performed on a variety of initial conditions
show that a mill in three dimensions will eventually acquire a
non-zero center-of-mass velocity and evolve into a flock as shown in
Fig.~\ref{fig:snapshots}(b).  In Fig.~\ref{fig:ztest9_time}(a), we
plot the state indicator $I_{\rm s}$ defined in the Appendix to characterize
the swarming pattern.  A value of $I_s = +1$ represents a perfect
unidirectional flock, $I_s=0$ a random collection of particles, and
$I_s = -1$ a perfect rotating mill. The red curve in
Fig.~\ref{fig:ztest9_time}(a) shows particles settling into a
transient mill for a lengthy period of time before evolving into a
translating flock; in contrast, the blue curve shows particles forming
a flock without first assembling into a long-lived mill. In the latter
case $I_{s}$ can first decrease before rising back to $I_{s} \approx
1$.

% 
%%%%%%%%%%%%%% Evolution without hydrodynamics %%%%%%%%%%%%%%%%%%%%%%%%%%%%%%%%%%
\begin{figure}[h!]
  \begin{center}
      \includegraphics[width=3.3in]{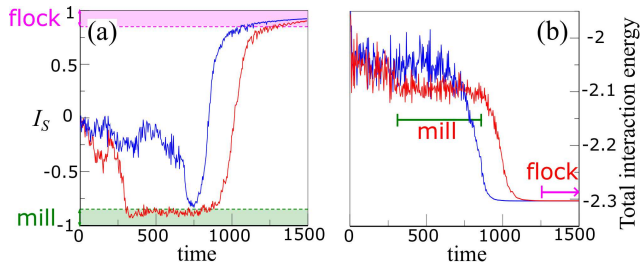}
\end{center}
  \caption{(a) Three-dimensional simulations of
    Eqs.~\ref{MDOT3} without hydrodynamic
    interactions. The classifier of swarming patterns $I_s$ is defined
    in the Appendix. The red curve denotes a swarm that first forms a mill
    before turning into a uniformly translating flock, while the blue
    curve shows particles evolving into a flock without first forming
    a mill.  We empirically set thresholds $I_s \le -0.85$ (green) to
    signal mills and $I_s \ge 0.85$ (magenta) to identify migrating
    flocks. We require a swarm to maintain the $I_s$ value in either
    the two ranges for a period of $100$ time units or more to be
    classified as a mill or flock. This criterion corresponds roughly
    to the time for a particle to circle a mill at least 10 times. (b)
    The total interaction energy $\sum_{i,j} \Phi \left( |\r_i - \r_j|
    \right) / 2$ corresponding to the two simulations above. The
    translating flock has a lower interaction energy than the mill.}
\label{fig:ztest9_time}
\end{figure}
%%%%%%%%%%%%%%%%%%%%%%%%%%%%%%%%%%%%%%%%%%%%%%%%%%%%%%%%%%%%%%%%%%%%%%%%%%%%
%

To understand how mills and flocks develop in three dimensions, in
Fig.~\ref{fig:ztest9_time}(b) we plot the evolution of the total
interaction energy ${1\over 2}\sum_{i,j} \Phi \left( |\r_i - \r_j|
\right)$ associated with the two simulations in
Fig.~\ref{fig:ztest9_time}(a), showing a lower total energy for the
flock state. Note that when assembled into flocks, particles settle
into positions that correspond to the \textit{global} minimum of the
total potential energy. In contrast, when assembled into mills, only a
\textit{local} minimum of the total potential energy is reached.  In
this case the net interaction force on each particle provides the
centripetal force necessary to sustain the rotational movement.
Although its energy is lower, for particles starting from random
initial conditions, a flock may be kinetically less accessible than a
mill.  A mill is a state of local coherence, where particles match
velocities with their close neighbors only, as opposed to a flock
where global coherence arises from all particles moving in unison.  As
a result, three-dimensional mills often emerge first out of a
randomized configuration. For the same reason, two-dimensional mills
are not only stable at steady-state, but also dominate over
flocks. However, three-dimensional mills are unstable since they
slowly acquire a translational momentum along the rotational axis
aligned with the third direction. As particle velocities gradually
become aligned with this translational momentum, the rotational unit
turns into a spiral with continuously reduced angular speed, finally
settling into an equilibrium lattice formation migrating at a uniform
velocity. This effect cannot arise in two dimensions.

\vspace{2mm}

\begin{figure}[ht]
  \begin{center}
      \includegraphics[width=3.4in]{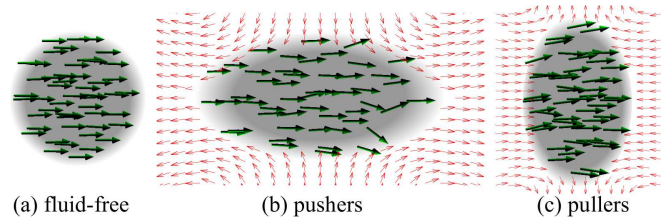}
\end{center}
\vspace{1mm}
  \caption{Snaphots of $50$-particle simulations showing
    the deformation of three-dimensional flocks by interacting with 
    $\u_{\rm s}$ arising in opaque fluids ($\lambda = 1$)
    when $\u_{\rm p}$ is absent. (a) For reference, we show the
    stable spherically shaped flock arising in the fluid-free
    case. This stable spherical flock is used as the initial conditions
    for both of the other two simulations before fluid interactions from 
    $\u_{\rm s}(\r)$ are switched on.
    (b) A transient prolate flock forms when $G=0.15$ (pusher).  (c) A
    transient oblate flock arises when $G=-0.15$ (puller). The red arrows 
    indicate the direction of $\u_{\rm s}$ in the frame of the flock
    center-of-mass.  In these simulations, the prolate flock is
    transitioning into a recirculating ``peloton,'' while the oblate
    flock is transitioning into a random blob.  These deformed flocks
    are stable only if $\vert G\vert$ is small, but with much less
    distortion. The emergence of pelotons and random blobs is
    described below.}
  \label{fig:deformed_flock}
\end{figure}

\noindent \textit{\textbf{Swimmer-induced fluid flow $\u_{\rm s}$ -- }}
We now investigate how patterns mediated by the $\u_{\rm s}$ flow
field alone differ from those described in the fluid-free case
above. In general, the extensional flow generated in the reference
frame of a puller ($G<0$) converges along the direction of motion and
diverges along the perpendicular direction.  Pusher-generated ($G>0$)
extensional flows move in the opposite direction, diverging along the
direction of motion and converging laterally. As a result, pullers
tend to flatten existing flocks into oblate shapes while pushers tend
to longitudinally stretch them into prolate shapes. These deformed
flocks are depicted in Fig.~\ref{fig:deformed_flock} for an opaque
fluid.

%%%%%%%%%%%%%%%%%%%%%%%%% prolate and oblate flocks  %%%%%%%%%%%%%%%%%%%%%%%%%%%
%%%%%%%%%%%%%%%%%%%%%%%%%%%%%%%%%%%%%%%%%%%%%%%%%%%%%%%%%%%%%%%%%%%%%%%%%%%%%%%

In clear fluids ($\lambda = 0$), the energy of a swarm dissipates
significantly through the fluid drag term, slowing particle motion and
reducing $\u_{\rm s}$.  For very large dimensionless drag $\gamma \gg
1$, the motion of both pushers and pullers is arrested and $\u_{\rm
  s}\to 0$. For intermediate $\gamma$, pushers align into prolate
flocks and move at a reduced speed of approximately $\sqrt{1 -
  \gamma}$; pullers also move at a reduced speed, but mostly randomly
without any spatial order. The $\gamma \to 0$ limit is the fluid-free
case.

\begin{figure}[ht]
  \begin{center}
      \includegraphics[width=3.3in]{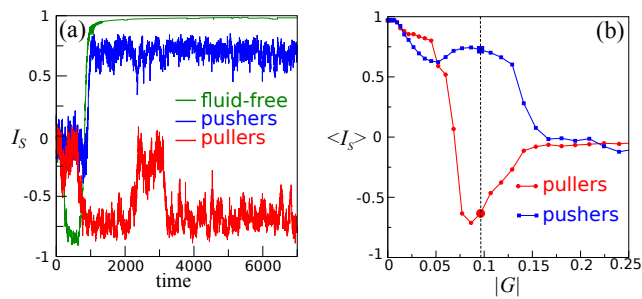}
\end{center}
  \caption{(a) Time-dependent swarm morphologies starting from random
    initial conditions.  Pusher swarms ($G=0.096$, blue curve), evolve
    directly towards a fluctuating flock, while puller swarms
    ($G=-0.096$, red curve) assemble into rotating mills; the latter
    persists indefinitely but is intermittently interrupted by bursts
    of randomness.  For comparison, we also plot the fluid-free case
    ($G=0$, green curve), where particles transiently form a mill
    before eventually assembling into a flock. (b) Long-time
    formations ($t>2000$) of pushers and pullers with different $G$
    values. The indicator $ \langle I_s \rangle $ is averaged for all
    time steps between $2000 \le t \le 3000$ and over ten simulations.
    For pullers, persistent mills only occur approximately in the
    range of $0.07 \leq G \leq 0.1$; below this range, flocks
    dominate, similarly to the fluid-free case; above this range, the
    swarm is in a permanent random state. Pushers always assemble into
    flocks; for large $G$, however, the flow field induces a
    peloton-like movement within the flock, which pushes $I_s$ towards
    zero.}
  \label{fig:puller_pusher}
\end{figure}

We observe a more diverse set of swarm morphologies in opaque fluids
($\lambda = 1$) where the self propulsion term $\f_{\rm M}$ imparts
sufficient energy to the particles to keep them moving at their
preferred self-propulsion speed relative to the background flow. In
this case, the oblate/prolate deformation of flocks is more pronounced
than in clear fluids.  In Fig.\,\ref{fig:puller_pusher}(a) we show the
time evolution of 50 particles for $\left| G \right| = 0.096$. The red
(blue) curves represent pullers (pushers). For reference we 
also plot the fluid-free case ($G=0$) in the green curve. For such 
small $G$, pusher-generated flows suppress the transient milling seen 
in the fluid-free case leading to a stable prolate flock.  However, 
unlike the fluid-free case, pusher-generated flocks are not perfect 
and $I_s \approx 0.75 < 1$. Here, the spatial-temporal variations 
in $\u_{\rm  s}$ impart fluctuations to the direction of particle 
movement, preventing the formation of a perfectly aligned
flock. Puller-generated flows, on the other hand, allow for the
formation of permanent mills: the mill-to-flock transition that occurs
in the fluid-free case is blocked by the fluctuating flow field
allowing mills to be long-lived.  

Since the flow disturbance may be considered as a form of noise, our
findings are consistent with previous reports of noise-induced
flock-to-mill transitions in three dimensions \cite{STR08,ROM12}. The
latter show hysteresis in swarm morphology as a function of the noise
amplitude, a feature which we also observe with pullers as the
thresholds in $G$ between a flock and random blob depend on whether
$G$ is increased or decreased.  In addition, we note that disturbances
induced by $\u_{\rm s}$ lead to particles occasionally deviating from
their circulating trajectories and to a striking intermittent
disintegration and re-assembling of mills, a phenomenon that has not
been reported in noise-induced flock-to-mill transitions.
%However, much the same as flocks of pushers, mills of pullers are constantly disturbed by $\u_{\rm s}$.
%Here, particles occasionally deviate from their circulating trajectories, leading to a striking intermittent disintegration and
%re-assembling of mills, a phenomenon that has not been previously reported.
In Fig.\,\ref{fig:puller_pusher}(b), we conduct
a more thorough investigation of long-time swarming patterns by
varying $\left| G \right|$ for pullers (blue) and pushers
(red). Pushers assemble into flocks as $G$ increases,
but patterns are increasingly disturbed by $\u_{\rm s}$, leading to
decreased $I_s$.

%%%%%%%%%%%%%%%%%%%%%%%%% pullers/pushers  %%%%%%%%%%%%%%%%%%%%%%%%%%%%%%%%%%%%%%
%%%%%%%%%%%%%%%%%%%%%%%%%%%%%%%%%%%%%%%%%%%%%%%%%%%%%%%%%%%%%%%%%%%%%%%%%%%%%%% 

\begin{figure}
  \begin{center}
\vspace{-3mm}
    \includegraphics[width=1.4in]{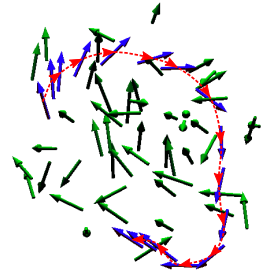}
\end{center}
  \caption{A snapshot of a peloton-like formation for $G=0.15$.
    Particle positions and velocities are represented by the green
    arrows, while the blue arrows connected by the red-dashed line
    track one particular particle for $15$ steps prior to the
    snapshot. The selected blue particle is initially near the leading
    edge of the flock, gets swept aside by the surrounding flow field,
    and finally rejoins the flock near the back end.}
  \label{fig:peloton}
\end{figure}

For larger $G$, $\u_{\rm s}$ is strong enough to induce a novel
``peloton''-like movement, where leading particles continuously
recirculate toward the back end of the flock, as depicted in
Fig.~\ref{fig:peloton}. When assembled into a peloton, $I_s$ drops to
nearly zero, although the majority of particles are still
aligned. Pullers on the other hand tend to keep milling as $\left| G
\right|$ increases rather than transition to a flock. Here $I_s
\approx -1$.  For very large values of $\left| G \right|$ the strong
flow field prevents even mills from forming, and particle movement
remains random.  Overall, our results suggest that pusher-generated
flow fields generally promote particle velocity ordering along a
common direction but that an orthogonal component of the flow prevents
perfect alignment for small $G $ and ultimately to particles
recirculating for large $G $. Puller-generated flow fields instead
introduce more randomness preventing the mill to flock transition for
small $\left| G \right|$ and completely preventing a mill from forming
for larger $\left| G \right|$.

%%%%%%%%%%%%%%%%%%%%%%%%% pullers/pushers  %%%%%%%%%%%%%%%%%%%%%%%%%%%%%%%%%%%%%%
%%%%%%%%%%%%%%%%%%%%%%%%%%%%%%%%%%%%%%%%%%%%%%%%%%%%%%%%%%%%%%%%%%%%%%%%%%%%%%% 

\vspace{2mm}

\noindent \textit{\textbf{Physical force-induced fluid flow
    $\u_{\rm p}$ --}} We now examine the effects of $\u_{\rm p}$ on
swarm dynamics by setting $G=0$ and analyze how patterns differ when
compared to those arising in the fluid-free case.  For a clear fluid
$(\lambda=0)$ our simulations reveal that at steady state particles
either stop or assemble into a flock. The resulting speed can be
evaluated by balancing self-propulsion with drag, yielding a
dimensionless flock speed $\sqrt{1 - \gamma}$ for $\gamma \le 1$ and
$0$ for $\gamma > 1$, which are both confirmed by simulations.  In
physical units, the friction threshold for immobilizing a flock is $6
\pi \eta a > \alpha$. Hydrodynamic coupling in a viscous clear fluid
simply slows or stops translational flock motion.  Note that in the
$a\to 0$ point-particle limit, drag is negligible and swarming in a
clear fluid reduces to the canonical fluid-free problem.

%%%%%%%%%%%%%% Transition Figure %%%%%%%%%%%%%%%%%%%%%%%%%%%%%%%%%%%%%%%%%%%%%%%
\begin{figure}[h!]
  \begin{center}
      \includegraphics[width=3.3in]{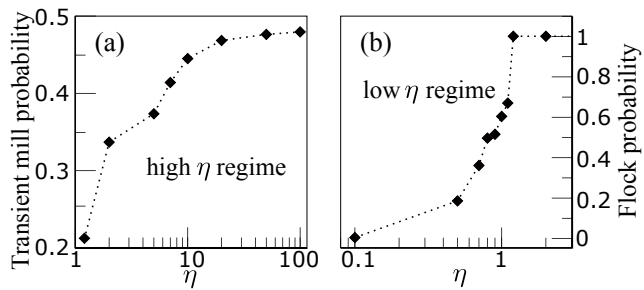}
\end{center}
  \caption{Swarm patterns in an opaque fluid as $\eta$ is varied.
    Each point is an average over 100 simulations each initialized
    with random conditions.  (a) Probability of transient mill
    formation for $\eta \ge 1.2$.  Although in this high viscosity
    regime all initial conditions lead to uniformly translating
    flocks, the probability of first forming a transient mill
    decreases with decreasing $\eta$. (b) Probability of permanent
    flock formation for $\eta \le 2$. In this low viscosity regime
    steady-state flocks are no longer the only final outcome when
    $\eta < 1.2$. Other possible configurations are mill-like
    structures and random blobs.}
\label{fig:ztest9}
\end{figure}
%%%%%%%%%%%%%%%%%%%%%%%%%%%%%%%%%%%%%%%%%%%%%%%%%%%%%%%%%%%%%%%%%%%%%%%%%%%%% 

As with swimmer-induced flows, the opaque fluid case is much more
interesting.  Here steady-state configurations depend nontrivially on
the dimensionless viscosity $\eta$ (defined in the Appendix) which measures
ratio of the fluid momentum relaxation time to the time scale of the
particle movement, thus representing an effective Deborah number for
the problem \cite{REINER1964,LAU09b}. The parameter $\eta$ appears in
Eq.~\ref{MDOT4} through $\u_{\rm p} (\r)$ in Eq.~\ref{USTATIC} and
through $\gamma = 6\pi \eta a$. We assume small particles and neglect
this latter drag interaction.  As can be seen from Eq.~\ref{USTATIC},
$\u_{\rm p}$ decreases with $\eta$ so that as $\eta$ increases the
dynamics resembles that of the fluid-free case.  Indeed, we find that
for $\eta > 1.2$, flocks are the only stable steady-state solution for
all random initial conditions used, similar to the fluid-free
case. However, transient mills can form before the permanent flock is
assembled, with the probability of transient mills occurring
decreasing with $\eta$.  As shown in Fig.~\ref{fig:ztest9}(a), for
$\eta = 100$ particles form mills before finally settling into flocks
for about $50\%$ of the random initial conditions used; this ratio
drops to about $20\%$ at $\eta = 1.2$.

\begin{figure}[htb]
  \begin{center}
      \includegraphics[width=2.9in]{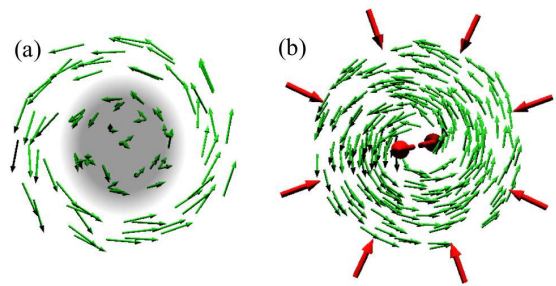}
\end{center}
  \caption{(a) A stable mill-like formation of 100 particles induced
    by hydrodynamic interactions in an opaque fluid with $\eta =
    0.7$. The gray shading delineates a disordered core that grows as
    $\eta$ is decreased.  (b) A transient mill of 250 particles with
    an empty core in an opaque fluid. Such transient mills may arise
    when $\eta > 1.2$. The induced flow field is indicated by the red
    arrows and converges toward the central core along the plane of
    rotation. To balance the influx, a jet along the axis of rotation
    ejects the fluid from the central core. The jet is normal to the page and depicted
    by red arrow in panel (b).}
  \label{fig:OseenField1000}
\end{figure}

\noindent Below $\eta \approx 1.2$, swarms experience a qualitative
change in behavior and flocks are no longer the only long-lived
steady-state.  Fig.~\ref{fig:ztest9}(b) shows that the probability of
final flock formation decreases from unity at $\eta \approx 1.2$ to
zero at $\eta \approx 0.1$. In this intermediate range of $\eta$, two
other long-lived configurations can arise: a mill-like formation as
shown in Fig.~\ref{fig:OseenField1000}(a), and a perpetual random
blob. Unlike the annular- or toroidal-shape of a classical mill, the
hydrodynamically-mediated three-dimensional mill-like structure has a
central core filled with randomly moving particles.  As the
dimensionless viscosity decreases from $\eta \approx 1.2$, the
randomly-moving core particles in a mill-like swarm expand their
boundaries and eventually swallow the coherent part of the mill. The
resulting pattern is a perpetual random blob without any identifiable
spatial order.  Finally, for $\eta \lesssim 0.1$, swarms immediately
collapse into the above described blob of perpetual random movement.
All possible swarming patterns are listed in Table 1 as a function of
the dimensionless viscosity $\eta$.  A viscous flow $\u_{\rm p}$ thus 
allows for the emergence of persistent mill-like
structures not observed in the absence of fluid flows.
%
%%%%%%%%%%%%%% Oseen Field 200 particles %%%%%%%%%%%%%%%%%%%%%%%%%%%%%%%%%%%%%%
%%%%%%%%%%%%%%%%%%%%%%%%%%%%%%%%%%%%%%%%%%%%%%%%%%%%%%%%%%%%%%%%%%%%%%%%%%%%%%% 
%
\begin{table}[htb]
%\centering
\begin{tabular}{|l|l|l|}
\hline
\backslashbox{$\eta$}{time} & intermediate $t$ & long $t$ \\
\hline
 $\eta > 1.2$ & mill or flock & flock only  \\ 
\hline
\: & mill or flock & flock \\
$0.1 < \eta < 1.2$ & mill-like or random & mill-like \\
\: & random & random \\ 
\hline
$\eta < 0.1$ & random & random \\
\hline
\end{tabular}
\begin{center}
\caption{Swarming structures observed in simulations for intermediate
  and long ($t > 3000$) times under the $\u_{\rm p}$ flow field and as
  a function of the dimensionless fluid viscosity $\eta$. Steady state
  configurations take longer to assemble here than under $\u_{\rm
    s}$. In the fluid-free case only flocks arise at long times. A
  moderately viscous fluid allows for the emergence of permanent
  mill-like structures and random blobs.}
\end{center}
\label{TABLE1}
\end{table}

\noindent
We can also examine the induced flow fields in relation to particle
velocities, and how they may drive transitions among various swarm
morphologies.  Starting from a high-$\eta$ transient mill regime
($\eta \gtrsim 1.2$), Fig.~\ref{fig:OseenField1000}(b) qualitatively
indicates the instantaneous direction of the hydrodynamic velocity
field $\u_{\rm p} (\r)$ (red arrows) induced by a mill-like formation
of $250$ particles.  In a transient mill, the net particle-particle
interactions provide the centripetal force that sustains
rotation. This net force is imparted on the fluid, inducing an inward
flow along the plane of rotation. The incompressible fluid is then
ejected outward along the rotational axis, resembling a ``jet''
emanating from the center of an accretion disk.
%
%%%%%%%%%%%%%%%%%%%%%%%%% Jet flow field  %%%%%%%%%%%%%%%%%%%%%%%%%%%%%%%%%%%%%%
\begin{figure}[ht]
  \begin{center}
      \includegraphics[width=3.4in]{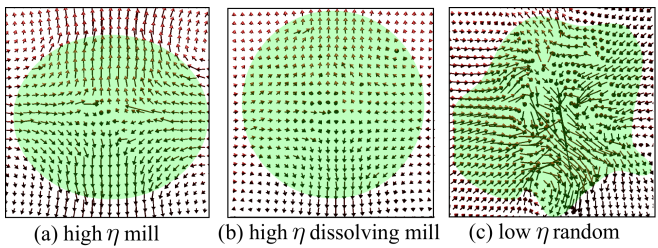}
\end{center}
  \caption{Fluid velocity fields $\u_{\rm p}(\r)$ associated with
    particle swarms concentrated within the green shaded regions. (a)
    For large $\eta$, an early flow field resembles a jet. (b) At
    longer times the jet is eventually disrupted and a flock
    forms. (c) For very low $\eta < 0.1$, the disordered core region
    of a transient mill-like formation will always expand eventually
    leading to a random blob.}
  \label{fig:jetflow}
\end{figure}
%%%%%%%%%%%%%%%%%%%%%%%%%%%%%%%%%%%%%%%%%%%%%%%%%%%%%%%%%%%%%%%%%%%%%%%%%%%%%%% 
% 
\noindent This outward jet entrains particles that wander into the
core region, slowly disrupting the mill. Entrainment arises through
the self-propulsion term $\f_{\rm M}$, and if
appreciable, through viscous drag.  Moreover, the inward flow on the
rotational plane effectively extends the interaction range among
particles, driving the system into a minimum-energy flock state.  As
$\eta$ is decreased, the induced Stokes accretion flow increases and drives
more particles into the core of the mill. Particle motion then becomes
randomized, disrupting the outward jet and ultimately
\textit{hindering} the mill-to-flock transition that would otherwise
occur smoothly.  Swarms can thus be trapped in the mill-like formation
shown in Fig.~\ref{fig:OseenField1000}(a) indefinitely as listed in
Table 1. At even lower values of $\eta$, coherence is lost by an
expanding disordered core region.  The fluid flow fields observed
under different regimes of $\eta$ are plotted in
Fig.~\ref{fig:jetflow}.

\vspace{2mm}

\noindent \textit{{\bf Combined effects of $\u_{\rm s}$ and $\u_{\rm
      p}$ -- }} In light of the above discussions, we now consider the
effects of superimposing the two fields so that $\u(\r) = \u_{\rm
  s}(\r) + \u_{\rm p}(\r)$. The magnitudes of $\u_{\rm s}$ and
$\u_{\rm p}$ can be varied independently, and are controlled by $G$
and $\eta$, respectively. In clear fluids ($\lambda = 0$), the two
flows combine to reduce flock speed to $\sqrt{1 - \gamma}$, except in
the case of very strong puller flows (large $|G|$) that prevent
particles from aligning into flocks and lead to a random blob.

\begin{figure}[ht]
  \begin{center}
      \includegraphics[width=2.7in]{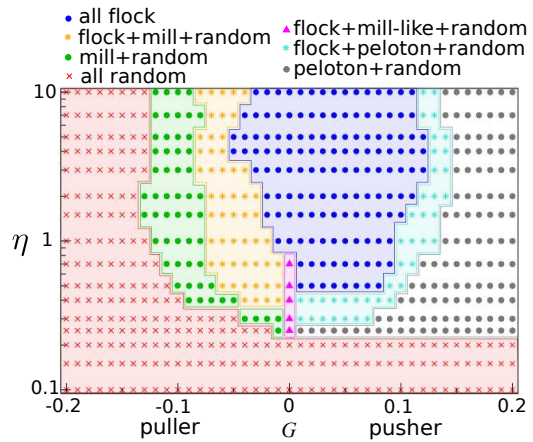}
\end{center}
  \caption{Phase diagram in $(G,\eta)$-space delineating possible
    persistent structures ($t > 3000$) in the presence of the total
    flow field $\u(\r) = \u_{\rm s}(\r)+ \u_{\rm p}(\r)$.  Each point
    summarizes the possible final morphologies from $10$
    simulations. Flocks are more likely to emerge for large $\eta$,
    and random blobs are more prominent for small $\eta$.  For pullers
    ($G<0$), mills may appear with increasing $\left| G \right|$, but
    random blobs dominate at large $\left| G \right|$.  The flow
    $\u_{\rm s}$ generated by weak pushers ($G\gtrsim 0$) promotes
    particle alignment and flock formation. For even larger $G$, flocks
    exhibit peloton-like movement.}
  \label{fig:combined_maps}
\end{figure}

Fig.\,\ref{fig:combined_maps} shows the phase diagram in 
($G,\eta$)-space of stable swarm structures arising in opaque fluids
$(\lambda = 1)$.  In opaque fluids for small values of 
$\eta$, $\u_{\rm p}$ dominates $\u_{\rm s}$ and swarms assemble into a 
random blob.
As $\eta$ increases, $\u_{\rm p}$ decreases and the effects of $\u_{\rm
  s}$ become more pronounced, prevailing for large $\eta$. In this
case, flows generated by strong pullers (very negative $G$) still
favor the emergence of a random blob.  However, upon increasing $G$
fluctuating transient mills arise.  As $G$ keeps increasing, flows
generated by pushers favor the formation of fluctuating flocks until
for very large $G$ pelotons emerge.
As can be seen in Fig.\,\ref{fig:combined_maps}, mill-like patterns
exist only when $G = 0$, suggesting that such structures are easily
disrupted or prevented from forming by $\u_{\rm s}$. 

%%%%%%%%%%%%%%%%%%%%%%%%% combined_maps  %%%%%%%%%%%%%%%%%%%%%%%%%%%%%%%%%%%%%%
%%%%%%%%%%%%%%%%%%%%%%%%%%%%%%%%%%%%%%%%%%%%%%%%%%%%%%%%%%%%%%%%%%%%%%%%%%%%%%% 
% 

\section*{Summary and Conclusions}

\noindent
In the absence of hydrodynamic coupling, our extensive numerical
simulations revealed that three-dimensional swarms exhibit much less
diversity than in two dimensions.  This is due to the additional
dimension that provides a pathway for a variety of stable
two-dimensional patterns to transform into energy-minimizing,
uniformly translating flocks in three dimensions.  We then carefully
explored the effects of hydrodynamic coupling on three-dimensional
swarming by deriving a discrete model of self-propelled interacting
swimmers in an incompressible zero-Reynolds-number Newtonian fluid.
 
An important distinction in the source of fluid flow is made. Under a
force-free assumption, particle swimming (or squirming \cite{ISH06})
can only generate flows that decay as $1/r^n, n\geq 2$.  When direct
action-at-a-distance physical forces (electrostatic, magnetic,
gravitational) arise between self-propelled particles, an additional
Oseen flow decaying as $1/r$ can arise \cite{HAP65,ESP95,BATCHELOR}.
Thus, even short-ranged pairwise physical forces can generate longer-ranged
fluid motion enhancing particle interactions and greatly affect swarm
morphology.

In clear three-dimensional fluid environments we find that only flocks
arise, similar to the fluid-free scenario, albeit with particles
moving at reduced speed. Note that our patterns emerge through direct
interactions among particles, in contrast to the ones observed in
previous studies of infinite or confined systems of swimmers coupled
only via the fluid drag \cite{BAS09}. The latter models include
particle density as a fixed, prescribed parameter, so that
high-density particles can be forced to interact out of equilibrium.
As a result, density-dependent transitions and states, such as nematic
orders, may arise.  On the contrary, we consider a finite number of
particles in an infinite domain, where local particle density is
determined by particles collectively minimizing the interactions
amongst them, favoring the low-energy flock formation.  In opaque
fluids, pusher generated flows accelerate particle alignment and
suppress the emergence of metastable mills seen in the fluid-free
case.  Puller-generated flows, conversely, hinder particle velocity
alignment, allowing transient mills to persist within certain
viscosity ranges. Sufficiently strong puller flows disrupt any spatial
order. Flows generated by particle-particle interactions kinetically
accelerate the mill-to-flock transition. In high-viscosity opaque
fluids, the hydrodynamic flow fields can form an accretion disk/jet
structure associated with mills and entrain the self-propelled
particles leading to quicker dissolution of the mill itself. However,
in opaque fluids of intermediate viscosity, stronger hydrodynamic
interactions may kinetically block the mill-to-flock transition,
allowing a mill-like formation to form and persist. At even lower
fluid viscosity $\eta$, swarms are completely chaotic.

When both swimming- and force-induced flows are
present, steady-state configurations depend on the relative strength
between the two flows.  Mill-like formations are absent.
Our main results pertain to viscous steady-state Stokes' flows, but we
derive time-dependent interaction-induced fluid velocities and
inertial interactions arising in potential flows in the Appendix.  We used
the Morse potential in our simulations to provide a mechanistic
picture of collective behavior of three-dimensional swarms and found a
rich phase diagram of patterns; however, the structure
of our fluid-coupled swarming model is sufficiently general that any
effective interaction potential can be used provided the its social or
physical underpinnings are carefully delineated. 

Our swarming model can be further refined by addressing more
microscopic fluid coupling mechanisms. In our model, Eq.\,\ref{eq:uv}
defines a stroke period-averaged flow field and Eq.\,\ref{MDOT3}
describes hydrodynamic coupling under the period-averaged flow
assumption. However, the phase difference of the microscopic strokes
between two swimmers has been shown to actively affect the interaction
in unexpected ways, leading to attraction, repulsion, or oscillations
\cite{POO07}.  How these subtle swimmer-induced pairwise interactions
\cite{ISH06,POO07} influence the collective behavior of swarms remains
to be investigated. Finally, we have not considered the effects of
external potentials. Our derivation of the fluid coupling
mechanisms suggest the possibility of more new structures depending on
whether the external potential derives from social interactions that
influence self-propulsion (\textit{e.g.}, chemotaxis) or physical ones
(\textit{e.g.}, gravity) that result in body forces on the fluid.  The
physical and mathematical structure of our fluid-coupled kinetic
models provide a basis for future investigation of these extensions.

\begin{acknowledgments}
The authors thank Eric Lauga and Lae Un Kim for helpful discussions.
This work was made possible by support from grants NSF DMS-1021818
(TC), NSF DMS-1021850 (MRD), ARO W1911NF-14-1-0472, and ARO MURI
W1911NF-11-10332. MRD also acknowledges insights gained at the
NAFKI conference on Collective Behavior: From Cells to Societies.
\end{acknowledgments}

\bibliographystyle{apsrev}
\bibliography{refs}

\newpage

%\section{SUPPLEMENTARY MATERIAL}
\appendix

\section{Non-dimensionalization}
\noindent
The non-dimensional parameters used in Eq.\,\ref{MDOT4} are defined as

\begin{equation}
   \begin{array}{lll}
    \r^\prime = \displaystyle{\frac{\sqrt{\alpha \beta}}{m} \r},  & t^\prime = \displaystyle{\frac{\alpha}{m} t,} \\  \\
      \v_i^\prime = \displaystyle{\sqrt{\frac{\beta}{\alpha}}} \v_i, & 
      \u^\prime = \displaystyle{\sqrt{\frac{\beta}{\alpha}}} \u, &
      \gamma^\prime = \displaystyle{\frac{\gamma}{\alpha}}, \\
\\
      \rho^\prime = \displaystyle{\frac{m^2}{\sqrt{\alpha^3 \beta^3}}} \rho, &
      \eta^\prime = \displaystyle{\frac{m}{\sqrt{\alpha^3 \beta}}} \eta, & 
      G^\prime = \displaystyle{\frac{\alpha \beta}{m^2}} G, \\
%      \nu^\prime = \frac{\beta}{m} \nu; \\
\\
      \Phi^\prime = \displaystyle{\frac{\beta}{\alpha m}} \Phi, & 
      p^\prime = \displaystyle{\frac{m^2}{\sqrt{\alpha^5 \beta}}} p, & \\
\\
      C_a^\prime = \displaystyle{\frac{\beta}{\alpha m}} C_a, & 
      C_r^\prime = \displaystyle{\frac{\beta}{\alpha m}} C_r, & \\ 
\\
      \ell_a^\prime = \displaystyle{\frac{\sqrt{\alpha \beta}}{m}} \ell_a, & 
      \ell_r^\prime = \displaystyle{\frac{\sqrt{\alpha \beta}}{m}} \ell_r.\tag{S8}
   \end{array}
\end{equation}

\section{Time-dependent Stokes flow \label{sec:time_dep_Stokes}}

We assume $\nu \rightarrow 0$ for the dimensionless Stokes
Eq.~\ref{MDOT4} of the main text and conduct our investigation at the
quasistatic limit. More generally, the time-dependent velocity field
can be expressed as

\begin{equation}
\u(\r_{j},t) = \frac {1} {\rho} \sum_{i\neq j}\int_{0}^{t}\dd t'{\bf
  T}(\r_{i}-\r_{j};t-t')\cdot\f_{i},\tag{S9}
\label{UOSEEN}
\end{equation}
where $\rho$ is the mass density of the embedding Newtonian fluid and
${\bf T}$ is the three-dimensional dynamic Oseen tensor given by \cite{ESP95}

\begin{equation}
\begin{array}{rl}
{\bf T}(\r, t) & \displaystyle = \int{\dd^{3}\k \over (2\pi)^{3}} e^{-\nu k^{2} t  + i\k\cdot\r}
\left[{\bf I} - \hat{\k}\hat{\k}\right] \\[13pt]
\: & \displaystyle = p(r,t){\bf I} - q(r,t)\hat{\r}\hat{\r},\tag{S10}
\label{OSEENT}
\end{array}
\end{equation}
with 
\begin{equation}
\begin{array}{l}
\displaystyle p(r,t) = \left(1+{2\nu t\over r^{2}}\right)f(r,t) - {g(r,t)\over r^{2}} \\[13pt]
\displaystyle q(r,t) = \left(1+{6\nu t\over r^{2}}\right)f(r,t) - {3g(r,t)\over r^{2}} \\[13pt]
\displaystyle f(r,t) = {1\over (4\pi \nu t)^{3/2}}\exp\left[-{r^{2}\over 4\nu t}\right] \\[13pt]
\displaystyle g(r,t) = {1\over 4\pi r} {\rm erf}\left({r \over \sqrt{4\nu t}}\right).\tag{S11}
\label{PQ}
\end{array}
\end{equation}
In the quasistatic limit, the Oseen tensor is

\begin{equation}
{1 \over \rho} {\bf T}(\r,t) \approx {1 \over 8\pi \eta r}\left[{\bf I}
  + \hat{\r}\hat{\r}\right]\delta(t), \tag{S12}
\label{OSEENSTATIC}
\end{equation}
which is used in Eq.~\ref{USTATIC} to provide an analytic form of the
velocity field. The solution to the pressure field $p(\r,t)$ is also
given; however, its effect on swimmers is ignored since its gradient
is negligible across the size $a \to 0$ of small particles.

\section{Potential flow \label{sec:potential_flow}}

\begin{figure}[ht]
  \begin{center}
      \includegraphics[width=3.2in]{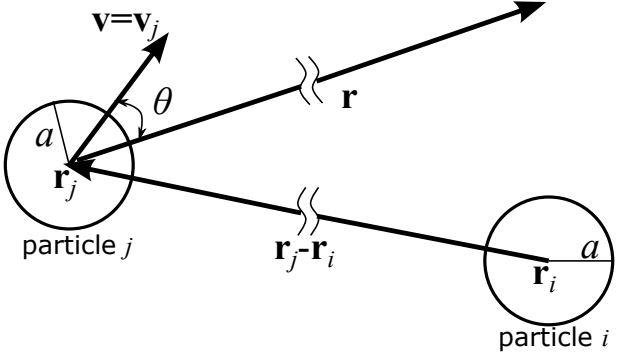}
\end{center}
  \caption{Hydrodynamic interactions due to potential flow.
    Eqs.~\ref{eq:far_poten00}-\ref{eq:far_poten01} express the
    approximated fluid potential at an arbitrary far-field location
    caused by the acceleration of a particle at $\r_j$.
    Eq.~\ref{eq:flow_potential_force} gives the force induced by the
    fluid potential on a particle at $\r_i$.}
  \label{fig:potential_flow}
\end{figure}

As derived in \cite{BLA88}, the fluid velocity potential $\phi(\r)$
at a location $\r$ from an accelerating spherical particle of radius $a$
can be approximated in the far-field $\vert \r\vert \gg a$
limit by the formula

\begin{equation}
   \phi \left(\r, t \right) \sim - \frac{m \left( t \right)}{4 \pi
     \left|\r \right|} - \frac{\mathbf{d} \left( t \right) \cdot
     \r}{4 \pi \left| \r \right|^3} + \mathcal{O} \left( \left| \r
   \right|^{-3} \right),\tag{S13}
\label{eq:far_poten00}
\end{equation}
where

\begin{equation}
   m \left(t \right) = \oint_{\partial V} \nabla \phi \cdot
   \mathrm{d} \mathbf{S}, \tag{S14}\label{eq:monopole}
\end{equation}
and

\begin{equation}
\mathbf{d} \left( t \right) = \oint_{\partial V} \phi \mathrm{d}
\mathbf{S} + \oint_{\partial V} \mathbf{r} \nabla \phi \cdot
\mathrm{d} \mathbf{S}\tag{S15} \label{eq:dipole}
\end{equation}
are obtained by integrating over the boundary $\partial V$ of the
spherical volume $V$ of the source object.  The near-field $\phi$ in
the integrands depends on the shape and swimming mechanism of the
source object.  Let us consider the simplest case of solid
spherical particles moving through an inviscid fluid.  For a lone
particle of radius $a$ traveling at a velocity $\v$ as illustrated in
Fig.~\ref{fig:potential_flow}, we may derive the fluid velocity
potential in the laboratory frame as follows
\begin{equation}
   \phi (\r) = - v \frac{a^3}{2 \left| \r \right| ^2} \cos \theta, \tag{S16}\label{eq:potential_phi}
\end{equation}
 where $v = \left| \v \right|$, $\r$ is the a spatial position relative 
to the center of the particle, and $\theta$ is the  angle between $\r$ 
and $\v$.
%%%%%%%%%%%%%% Potential Flow Illustration %%%%%%%%%%%%%%%%%%%%%%%%%%%%%%%%%%%%%%
%%%%%%%%%%%%%%%%%%%%%%%%%%%%%%%%%%%%%%%%%%%%%%%%%%%%%%%%%%%%%%%%%%%%%%%%%%%%%%%
\noindent Substituting Eq.~\ref{eq:potential_phi} into Eqs.~\ref{eq:monopole}
and \ref{eq:dipole}, we obtain
\begin{equation*}
   \begin{array}{l}
      m \left( t \right) = 0,\quad\mbox{and}\quad \mathbf{d} \left( t
      \right) = 2 \pi a^3 \v \left( t \right).
   \end{array}
\end{equation*}
As a result, the far-field fluid velocity potential of the moving sphere is

\begin{equation}
   \phi \left(\r, t \right) \sim - \frac{a^3 \v \left( t \right)
     \cdot \r}{2 \left| \r \right|^3},\tag{S16}
         \label{eq:far_poten01}
\end{equation}

\noindent Let us now calculate the force induced by a moving particle at 
position $\r_{j}$ on an identical particle at a position $\r_i$.
We assume that $\left| \r_i - \r_j \right| \gg a$, so that the
far-field approximation is appropriate.  From the Euler equation of
inviscid flow, we know that the fluid velocity potential $\phi$
induces a pressure
\begin{equation}
p = p_0 - \rho \frac{\partial \phi}{\partial t} - \frac{1}{2} \rho
\left| \u \right|^2,\tag{S17}
\label{eq:potential_pressure}
\end{equation}
where $\rho$ is the fluid density, $\u = \nabla \phi$ is the fluid 
velocity, and $p_0$ is an arbitrary reference point of the pressure.
The resultant force on a spherical object is
found from integrating the pressure variation over its surface:

\begin{align}
   \F_\mathrm{p.f.} & \displaystyle = \oint_{\partial V} p \mathrm{d} 
     \mathbf{S} \nonumber \\[13pt]
\: & \displaystyle = - \oint_{\partial V} \rho \frac{\partial
     \phi}{\partial t} \mathrm{d} \mathbf{S} \nonumber \\[13pt]
\: & \displaystyle  = \frac{\rho}{4
     \pi} \frac{\mathrm{d} \mathbf{d} \left( t \right)}{\mathrm{d} t}
   \cdot \oint_{\partial V} \frac{\left( \r - \mathbf{s}
     \right)}{\left| \r - \mathbf{s} \right|^3} \hat{\mathbf{s}}
   \mathrm{d} S.\tag{S18}
\label{eq:pressure_force}
\end{align}
Here, $\r\equiv \r_j - \r_i$, $\mathbf{s}$ is a vector from the
particle center to the particle surface, and $\hat{\mathbf{s}} \equiv
\mathbf{s} / \left| \mathbf{s} \right|$.  For a spherical particle of
radius $a = \left| \mathbf{s} \right| \ll \left| \r \right|$, we
use the approximation $\left| \r - \mathbf{s} \right|^{-3} \simeq
\left| \r \right|^{-3} \left( 1 + 3 \left( \mathbf{s} \cdot \r
\right)/\left| \r \right|^2 \right)$ to find

\begin{equation}
   \F_\mathrm{p.f.} = \frac{\rho a^3}{3 \left| \r \right|^3}
   \dot{\mathbf{d}}(t)\cdot \left[ 3
     \frac{\r \r}{\left| \r \right|^2} - \mathbf{I} \right].\tag{S19}
\end{equation}
Substituting Eq.~\ref{eq:dipole} into the above equation, we obtain

\begin{equation}
   \F_\mathrm{p.f.} = \frac{2 \pi \rho a^6}{3 \left| \r \right|^3}
   \dot{\v}(t) \cdot \left[ 3 \frac{\r
       \r}{\left| \r \right|^2} - \mathbf{I} \right].\tag{S20}
\end{equation}

\noindent Assuming there are $N$ identical particles, the potential flow induced
force on particle $i$ is thus
\begin{equation}
   \F^{i}_\mathrm{p.f.} = \frac{2 \pi \rho a^6}{3} \sum_{j \ne i}^N
   \frac{\dot{\v}_{j}(t)}{\left|\r_j - \r_i \right|^3} 
\cdot \left[ 3 \frac{\left( \r_j - \r_i \right)
       \left( \r_j - \r_i \right)}{\left| \r_j - \r_i \right|^2} -
     \mathbf{I} \right].\tag{S21}
             \label{eq:flow_potential_force}
\end{equation}
Note that the force is short-ranged, of the order $\mathcal{O} \left(
\left| \r_j - \r_i \right|^{-3} \right)$.  Moreover, its amplitude is
proportional to $a^{6}$. As a result, the hydrodynamic interaction
force induced by potential flows does not have significant impact on
collective behavior particularly when $\rho$ and/or the volume
fraction of particles is small. If $\rho$ is very large, the flow
field imposes a repulsion between particles that encounter each other,
potentially preventing them from forming a coherent structure.

\section{Indicator of the swarming states \label{sec:indicator}}

Here we define a metric to describe the state of a swarm.  This
quantity will consistently distinguish between parallel flock, single
rotating mill, and random swarms.  To identify the parallel flock
state, we note that all particles are moving at the same velocity as
the Center-of-Mass (CM) velocity.  To find the rotating mill state, we
take advantage of the fact that all particles share the same axis of
rotation.  We combine these properties into a single quantity $I_{s}$
over the desired range $[-1, 1]$ where $-1$ is associated with a
perfect mill and $+1$ indicates a uniformly translating flock. The
indicator $I_{s}$ is decomposed according to

\begin{equation}
   I_s \equiv I_\mathrm{flock} - I_\mathrm{mill}. \tag{S22}\label{eq:Is}
\end{equation}
Given $N$ particles,

\begin{equation}
I_\mathrm{flock} \equiv 1 - \frac{\sum_i \left| \mathbf{v}_i
  -\mathbf{v}_\mathrm{CM} \right|}{N \sqrt{\alpha /
    \beta}}. \tag{S23}\label{eq:Ipf}
\end{equation}
Note that $I_\mathrm{flock} = 1$ for a perfect parallel flock and
$I_\mathrm{flock} = 0$ for a perfect mill.  To define
$I_\mathrm{mill}$ we first compute the rotational axis
$\hat{\mathbf{\omega}}_i$ of particle $i$:

\begin{equation}
\hat{\mathbf{\omega}}_i \left( t \right)  =  \frac{\mathbf{v}_i
  \left( t \right) \times \mathbf{F}_i \left(t \right)}{ \left|
  \mathbf{v}_i \left( t \right) \right| \left| \mathbf{F}_i \left( t
  \right) \right|}. \tag{S24}\label{eq:unit_R2}
\end{equation}
where $\mathbf{F}_i$ is the force acting on particle $i$.  We then
evaluate the degree of alignment between all $\hat{\mathbf{\omega}}_i$
and define

\begin{equation}
I_\mathrm{mill}  =  \frac{\sum_i \sum_{j \ne i}
  \hat{\mathbf{\omega}}_i \cdot \hat{\mathbf{\omega}}_j}{N \left( N -
  1 \right)}.\tag{S25} \label{eq:Icr}
\end{equation}
Note that $I_\mathrm{mill} = 1$ when the rotations of all the
particles are perfectly aligned and $I_\mathrm{mill} = 0$ when all
particles are in a perfect parallel flock formation.  Putting
$I_\mathrm{flock}$ and $I_\mathrm{mill}$ together in $I_{s}$
(Eq.~\ref{eq:Is}), we find $I_s = -1$ for a perfect mill and $I_{s}=
+1$ for a perfect flock. Finally, since swarms are seldom in a perfect
formation, we considered thresholds on $I_{s}$ as indicated in
Fig.~\ref{fig:ztest9_time}.

Distinguishing more subtly different structures is not always
unequivocal using the metric $I_{s}$. In particular, we prescribe $I_s
> 0.5$ to indicate a flock and $I_s < 0.5$ to indicate a peloton where
there is more rotational movement from particle recirculation.

\section{Effects of changing interaction potentials}

\begin{figure}[ht]
  \begin{center}
      \includegraphics[width=2.4in]{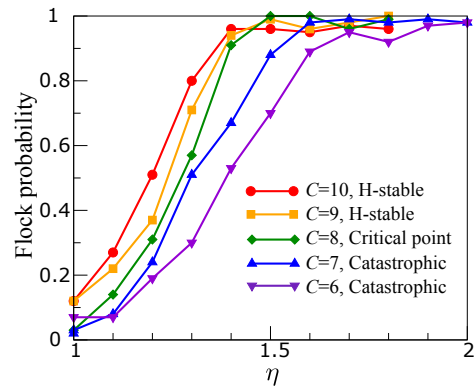}
\end{center}
  \caption{The dependence of stable flock formation probability on
    potential types and opaque medium viscosity. Here, we fix $\ell_a
    = 1$, $\ell_r=0.5$ and vary $C\equiv C_{r}/C_{a} = 6-10$ while
    keeping $C_{a} = 100$. H-stable flocks are more robust against
    hydrodynamic disruption.}
\label{fig:zsize15}
\end{figure}

All the results presented in the main text were obtained using a fixed
set of potential parameters $C_{r,a}, \ell_{r,a}$. The primary effect
of varying these parameters is to change the spatial size of swarms.
For rotational mills, an increase in diameter is accompanied by a
decrease in the magnitude of the centripetal force and weaker
destabilizing flows. A larger swarm is also less sensitive to
hydrodynamic effects since particles are spaced further apart,
generating weaker interaction forces and hence weaker flows.  In
Fig.~\ref{fig:zsize15}, we explore different potentials and test the
robustness of flock formation in the low $\eta$ regime where the flock
can be broken up by hydrodynamic interactions.  Not surprisingly, for
potentials that are more ``H-stable'' \cite{DOR06,CHU07b}, the
probability of stable flock formation increases. While H-stability is
an equilibrium property that is insensitive to hydrodynamics
\cite{DOR06,CHU07b}, our results suggest that H-stable flocks are more
resistant to hydrodynamic disruption.

%%%%%%%%%%%%%% Transition Figure %%%%%%%%%%%%%%%%%%%%%%%%%%%%%%%%%%%%%%%%%%%%%%%
%%%%%%%%%%%%%%%%%%%%%%%%%%%%%%%%%%%%%%%%%%%%%%%%%%%%%%%%%%%%%%%%%%%%%%%%%%%%%

%\end{article}  %comment out for revtex
\end{document}